\documentclass[%
 reprint,onecolumn,
superscriptaddress,
nofootinbib,
 amsmath,amssymb,
 aps,
]{revtex4}

\usepackage{graphicx}
\usepackage{dcolumn}
\usepackage{bm}
\usepackage{xcolor}
\usepackage{hyperref}
\usepackage{cleveref}

\begin{document}

\newtheorem{theorem}{Theorem}
\newtheorem{definition}{Definition}
\newtheorem{lemma}{Lemma}
\newtheorem{proposition}{Proposition}
\newtheorem{remark}{Remark}
\newtheorem{corollary}{Corollary}
\newtheorem{example}{Example}


\title{Anomalous and ultraslow diffusion of a particle driven by
power-law-correlated and distributed-order noises}

\author{Z. Tomovski}
\affiliation{University of Ostrava, Faculty of Sciences, Department of Mathematics, 30. Dubna 22701 03 Ostrava, Czech Republic}
\email{zhivorad.tomovski@osu.cz}

\author{K.~G\'{o}rska}
\email{katarzyna.gorska@ifj.edu.pl}
\affiliation{Institute of Nuclear Physics, Polish Academy of Sciences, ul.Eljasza-Radzikowskiego 152, PL 31342 Krakow, Poland}

\author{T.~Pietrzak}
\email{tobiasz.pietrzak@ifj.edu.pl}
\affiliation{Institute of Nuclear Physics, Polish Academy of Sciences, ul.Eljasza-Radzikowskiego 152, PL 31342 Krakow, Poland}

\author{R. Metzler}
\affiliation{Institute of Physics and Astronomy, University of Potsdam, D-14476 Potsdam-Golm, Germany}
\email{rmetzler@uni-potsdam.de} 
\affiliation{Asia Pacific Center for Theoretical Physics, Pohang 37673, Republic of Korea}

\author{T. Sandev}
\email{trifce.sandev@manu.edu.mk} 
\affiliation{Research Center for Computer Science and Information Technologies, Macedonian Academy of Sciences and Arts, Bul. Krste Misirkov 2, 1000 Skopje, Macedonia} 
\affiliation{Institute of Physics, Faculty of Natural Sciences and Mathematics, Ss. Cyril and Methodius University, Arhimedova 3, 1000 Skopje, Macedonia}
\affiliation{Institute of Physics and Astronomy, University of Potsdam, D-14476 Potsdam-Golm, Germany}

\begin{abstract}
We study the generalized Langevin equation approach to anomalous diffusion
for a harmonic oscillator and a free particle driven by different forms of
internal noises, such as power-law-correlated and distributed-order noises
that fulfil generalized versions of the fluctuation-dissipation theorem.
The mean squared displacement and the normalized displacement correlation
function are derived for the different forms of the friction memory kernel.
The corresponding overdamped generalized Langevin equations for these
cases are also investigated. It is shown that such models can be used to
describe anomalous diffusion in complex media, giving rise to subdiffusion,
superdiffusion, ultraslow diffusion, strong anomaly, and other complex
diffusive behaviors.
\end{abstract}

\date{\today}

\maketitle

\section{Introduction}

The behavior of a test particle of mass $m$, that is coupled to a thermal bath
of temperature $T$, can be described by Newton's second law for a particle in
the presence of a deterministic external potential $V(x)$ and a stochastically
varying force $\xi(t)$. When the friction acting on the test particle is given
by the constant value $\gamma_0$, the resulting dynamic is described by the
standard Langevin equation for a Brownian particle \cite{coffey,langevin,Zwanzig},
\begin{equation}
\label{lang}
m\ddot{x}(t)+m\gamma_0\dot{x}(t)+\frac{dV(x)}{dx}=\xi(t),
\quad \dot{x}(t)=v(t),
\end{equation}
where $x(t)$ is the particle displacement and $v(t)$ is its velocity. The
stochastic force, i.e., the "noise", $\xi(t)$ is Gaussian, has zero mean, and
is white such that its autocovariance is $\delta$-correlated, $\langle\xi(t+t')
\xi(t)\rangle=mk_BT\gamma_0$ with thermal energy $k_BT$. The mean-squared
displacement (MSD) encoded in the stochastic equation (\ref{lang}) scales
quadratically ("ballistically") at short times, at which it is dominated by
inertial effects, and then crosses over to a linear time dependence beyond the
characteristic time scale $1/\gamma_0$.

In the following we will consider generalizations of the stochastic equation
(\ref{lang}) that include "memory", by which we mean non-localities in time.
A simple example giving rise to memory is the Brownian harmonic oscillator
("Ornstein-Uhlenbeck process")
\begin{equation}
\label{oup}
\dot{x}(t)=v(t),\quad m\dot{v}=-m\omega^2 x(t)-m\gamma_0v(t)+\xi_v(t),
\end{equation}
where $\xi_v(t)$ is zero-mean white Gaussian noise. The subscript $v$ denotes
that this noise appears in the equation for $\dot{v}(t)$. On purpose, we here
use the phase space notation explicitly keeping the velocity $v(t)$ in the
second equation. Suppose that $v(0)=0$, so that we can integrate the second
equation from $t=0$, yielding
\begin{equation}
v(t)=\int_0^t e^{-\gamma_0(t-t')}\left(-\omega^2 x(t')+\xi_v(t')/m\right)dt'.
\end{equation}
Substituting this result back into the first equation in (\ref{oup}), we find
\begin{equation}
\label{ou-nm}
\dot{x}(t)=-\int_0^tK(t')x(t-t')dt'+\xi_x(t),
\end{equation}
where we defined the memory kernel $K(t)$ and the positional friction $\xi_x
(t)$,
\begin{equation}
K(t)=\omega^2e^{-\gamma_0t},\quad \xi_x(t)=\frac{1}{m}\int_0^te^{-\gamma_0t'}
\xi_v(t-t')dt'.
\end{equation}
At sufficiently long times $t\ll1/\gamma_0$ (or when we start the process at
$t=-\infty$) the Ornstein-Uhlenbeck process reaches equilibium and satisfies
the fluctuation-dissipation theorem $\langle\xi_x(t)\xi_x(t')\rangle\sim\langle
x^2\rangle_{\mathrm{eq}}K(|t-t'|)$ with the thermal value $\langle x^2\rangle_{
\mathrm{eq}}=k_BT/(m\omega^2)$.

From our result (\ref{ou-nm}), which is characterized by an exponential memory,
one thus cannot conclude that the underlying Ornstein-Uhlenbeck process is
non-Markovian. However, we showed a general property: namely, when integrating
out Markovian degrees of freedom, memory effects in the resulting equation for
the test particle of interest emerge \cite{Zwanzig}. Indeed, much more
pronounced memories, such as power-law forms, can be effected by eliminating
a quasi-continuum of Markovian degrees of freedom, see, e.g., \cite{goychuk,
kupfer,lizana,Zwanzig}.

For a general friction memory kernel $\gamma(t)$ the dynamics of a stochastic
process is described in terms of the generalized Langevin equation (GLE)
\cite{Zwanzig}
\begin{equation}
\label{GLE}
m\ddot{x}(t)+m\int_0^t\gamma(t-t')\dot{x}(t')dt'+\frac{dV(x)}{dx}=\xi(t),\quad \dot{x}(t)=v(t).
\end{equation}
The second fluctuation-dissipation theorem (FDT) is then valid in a thermal bath
of temperature $T$, where fluctuations and dissipation come from the same source.
The FDT relates the friction memory kernel $\gamma(t)$ with the correlation
function $\xi(t)$ of the random force \cite{kubo,Mainardi et al,Zwanzig}. The
FDT allows one to write
\begin{equation}
\label{correlationSFDT}
\langle\xi(t+t')\xi(t')\rangle=C(t)=k_BT\gamma(t).
\end{equation}
Here the friction memory kernel is assumed to satisfy \cite{Desposito_Vinales}
\begin{equation}
\label{assumption}
\lim_{t\rightarrow\infty}\gamma(t)=\lim_{s\rightarrow0}s\hat{\gamma}(s)=0,
\end{equation}
where the hat denotes the Laplace transform of $\gamma(t)$, i.e., $\hat{\gamma}
(s)=\mathcal{L}\{\gamma(t);s\}=\int_0^{\infty}\gamma(t)e^{-st}dt$.
We note that when the noise is external in the sense of Klimontovich \cite{klimo},
i.e., the noise is not provided by a heat bath in a non-equilibrium systems, the
relation \eqref{correlationSFDT} does not hold and $\langle\xi(t)\xi(t')\rangle=
C(t-t')$ is used instead. When we again consider the harmonic oscillator and use
the unit mass $m=1$ (i.e., $V(x)=\omega^2 x^2/2$), then \eqref{GLE} can rewritten
in the form \cite{Vinales_Desposito}
\begin{equation}
\label{X(t)}
x(t)=\langle x(t)\rangle+\int_0^tG(t-t')\xi(t')\mathrm{d}t', \quad v(t)=\langle
v(t)\rangle+\int_0^tg(t-t')\xi(t')\mathrm{d}t',
\end{equation}
where 
\begin{equation}
\label{X_av}
\langle x(t)\rangle=x_0\left[1-\omega^2I(t)\right]+v_0G(t),\quad \langle v(t)
\rangle=v_0g(t)-\omega^2x_0G(t)
\end{equation}
are the average particle displacement velocity, respectively, given the initial
conditions $x_0=x(0)$ and $v_0=v(0)$. Moreover, we introduced the so-called
relaxation functions $G(t)$, $I(t)=\int_0^tG(\xi)d\xi$ and $g(t)=dG(t)/(dt)$,
which in the Laplace space read
\begin{equation}
\label{G(s)}
\hat{G}(s)=\frac{1}{s^2+s\hat{\gamma}(s)+\omega^2},\quad\hat{g}(s)=s\hat{G}(s),
\quad\mbox{and}\quad\hat{I}(s)=s^{-1}\hat{G}(s).
\end{equation}
These functions are used to calculate the following four fundamental quantities
(i) the MSD \cite{Desposito_Vinales}
\begin{equation}
\label{19/05/23-1}
\langle x^2(t)\rangle=2k_BTI(t)=2k_BT\int_0^t G(\xi)d\xi;
\end{equation}
(ii) the diffusion coefficient ${\cal D}(t)=(1/2)[d\langle x^2(t)\rangle/(d
t)]$ that, due to relation \eqref{19/05/23-1} can be expressed as
\begin{equation}
\label{19/05/23-2}
{\cal D}(t)=k_BTG(t)=k_BT\frac{d}{dt} I(t),
\end{equation}
compare the proof of relation \eqref{19/05/23-2} for the GLE in the free-force
case in \cite{Mainardi et al, Pottier};\\
(iii) the normalized displacement autocorrelation function (DACF), that can be
experimentally measured which under the initial conditions $\langle x_0^2\rangle
=k_BT/\omega^2$, $\langle x_0v_0\rangle=0$, and $\left\langle\xi(t)x_0\right
\rangle=0$, can be represented as \cite{sandev_physmaced,sandev_jmp}
\begin{equation}
\label{NDCF thermal}
C_X=\frac{\langle x(t)x_0\rangle}{\langle x^2_0\rangle}=1-\omega^2I(t)=1-\omega^2
\int_0^tG(\xi)d\xi.
\end{equation}
(Note that different DACFs for fractional GLE are studied in \cite{sandev_jmp}.)\\
(iv) the normalized velocity autocorrelation function (VACF) \cite{Desposito_Vinales}
\begin{equation}
\label{19/05/23-5}
C_V(t)=\frac{\langle v(t)v_0\rangle}{\langle v^2_0\rangle}=g(t)=\frac{d}{dt}
G(t)=\frac{d^{2}}{d t^2} I(t).
\end{equation}

In the large friction limit, we neglect the inertial term $m\ddot{x}(t)$, and the
resulting overdamped GLE \eqref{GLE} has the form 
\begin{eqnarray}
\label{GLE overdamped}
\int_0^t\gamma(t-t')\dot{x}(t')dt'+\frac{dV(x)}{dx}=
\xi(t),\quad\dot{x}(t)=v(t).
\end{eqnarray}
The solution of this overdamped equation is of particular interest due to its
application in modeling the anomalous dynamics of colloidal (micron-sized) test
particles or the longer-time internal motion of proteins. Large friction, which
appears due to the liquid environment means that the acceleration $\ddot{x}(t)$
is negligible in comparison to the effect of the friction term. Single-particle
tracking of colloidal particles in an optical tweezers trap \cite{franosch,jeon,
jeon1,norre,selh} or the internal motion of proteins can be considered as the
effective motion in an harmonic potential \cite{Kou_Xie,yang}, the motion is
described by the GLE 
\begin{eqnarray}\label{GLE overdamped harmonic}
\int_0^t\gamma(t-t')\dot{x}(t')\mathrm{d}t'+m\omega^{2}x(t)=\xi(t),\quad \dot{x}
(t)=v(t).
\end{eqnarray}
The relaxation functions in the overdamped limit read
\begin{align}
\label{12/06/23-1}
G_o(t)=\mathcal{L}^{-1}\Big[\frac{1}{s\hat{\gamma}(s)+\omega^2};t\Big],\quad
g_o(t)=\mathcal{L}^{-1}\Big[\frac{s}{s\hat{\gamma}(s)+\omega^{2}};t\Big],\quad
\text{and}\quad I_o(t)=\mathcal{L}^{-1}\Big[\frac{s^{-1}}{s\hat{\gamma}(s)+\omega^{2}};t\Big].
\end{align}

For the case of a white Gaussian form for the noise $\xi(t)$ the GLE (\ref{GLE})
corresponds to the classical overdamped Ornstein-Uhlenbeck process equation with friction coefficient $\gamma_0$. In the force-free case this
further reduces to the Langevin equation (\ref{lang}) for a force-free Brownian
particle. At times $t\ll1/\gamma_0$ the MSD is then given by $\langle x^2(t)
\rangle\sim2[k_BT/(\gamma_0m)]t$, and thus the diffusion coefficient becomes
$\mathcal{D}=\lim_{t\to\infty}\langle x^2(t)\rangle/(2t)=k_BT/(m\gamma_0)$,
whose physical dimensions are $[\mathcal{D}]=\mathrm{length}^2/\mathrm{time}$.
The latter result for the diffusion coefficient of a Brownian particle in fact
represents the Einstein-Smoluchowski-Sutherland relation \cite{einstein,smolu,
spiecho,suth}.

Different forms for the friction memory kernel, particularly power-law forms
\cite{Burov_Barkai,Desposito_Vinales,Lutz,Mainardi et al,sandev_tomovski_pla,Wang}
and Mittag-Leffler (ML) forms \cite{sandev physa,sandev3,sandev_physmaced,
sandev_jmp,Vinales_Desposito} have been introduced to model anomalous diffusion,
for which the MSD scales non-linearly in time,
\begin{equation}
\label{msd}
\langle x^2(t)\rangle=\frac{2\mathcal{D}_{\alpha}}{\Gamma(1+\alpha)}t^{\alpha},
\end{equation}
where $\mathcal{D}_{\alpha}$ is the generalized diffusion coefficient with
physical dimension $\left[\mathcal{D}_{\alpha}\right]=\mathrm{length}^2/
\mathrm{time}^{\alpha}$, and where $\alpha$ is the anomalous diffusion
exponent. We distinguish the cases of subdiffusion ($0<\alpha<1$) and
superdiffusion ($1<\alpha$) \cite{Metzler3}. Anomalous diffusion of this
power-law form occurs in a multitude of systems across many scales
\cite{pt,cas,hoefling,pt1,igor_wings,vilk} and it is non-universal in the sense
that the MSD (\ref{msd}) for a given $\alpha$ may emerge from a range of
different anomalous stochastic processes \cite{stasrev,pccp,andi,henrik,igor_sm}.
We also mention that anomalous diffusion based on the GLE with crossovers to
a different $\alpha$ exponent or normal diffusion can be modeled in terms of
tempered power-law kernels as introduced in \cite{daniel} and applied to the
anomalous diffusion of lipids in bilayer membranes \cite{jeon_prl}. Similar
crossovers can be achieved in terms of formulations with distributed-order
kernels as discussed below. We also note that non-Gaussian processes with
power-law correlated noise were shown to emerge from a superstatistical
approach based on the GLE \cite{jakub,straeten} and interactions of GLE
dynamics with reflecting boundaries were analyzed in \cite{vojta}.

In previous work \cite{sandev_tomovski_pla} we considered the GLE for a free
particle driven by a mixture of $N$ independent internal white Gausian noises
\begin{equation}
\label{sum}
\xi(t)=\sum_{i=1}^N\alpha_i\,\xi_i(t),
\end{equation}
such that each has zero mean, $\langle
\xi_i(t)\xi_j(t')\rangle=0$ and correlation
\begin{eqnarray}
\label{correlation_i}
\langle \xi_i(t)\xi_i(t')\rangle=\delta_{ij}\,\zeta_i(t'-t),
\end{eqnarray}
where $\delta_{ij}$ is the Kronecker-$\delta$. The correlation function of
the additive noise $\xi(t)$ is then \cite{sandev_tomovski_pla}
\begin{eqnarray}
\label{correlation_final}
\langle \xi(t)\xi(t')\rangle=\left<\sum_{i=1}^N\alpha_i\,\xi_i(t)\sum_{j=1}^N
\alpha_j\,\xi_j(t')\right>=\sum_{i=1}^N\alpha_i^2\,\langle\xi_i(t)\xi_i(t')\rangle.
\end{eqnarray}
From the second FDT (\ref{correlationSFDT}) we thus see that the noise
fulfils
\begin{eqnarray}
\label{SFDT_final}
\sum_{i=1}^N\alpha_i^2\,\zeta_i(t)=k_BT\gamma(t),
\end{eqnarray}
where $\gamma(t)$ is the associated friction memory kernel. In
\cite{sandev_tomovski_pla} the GLE with internal noises of Dirac-$\delta$,
power-law, and ML types were analyzed and various different diffusive
regimes obtained. Moreover, it was shown that friction memory kernels of
distributed order can be used to describe ultraslow diffusion with a
logarithmic time dependence of the MSD. In what follows we consider a
stochastic harmonic oscillator driven by a mixture of internal noises,
from which we recover the results for a free particle in the limit of
vanishing force constant.

Here we analyze the MSD and DACF for different forms of the friction
memory kernel.  In Section \ref{sect2} we specify the GLE approach
to anomalous diffusion. We analyze the relaxation functions, MSD, and
DACF for Dirac-$\delta$, power-law, and combinations friction kernels.
The corresponding overdamped limits are analyzed and the force-free limit
recovered. Distributed-order friction memory kernels are considered in Section
\ref{sect3}. It is shown that such kernels yield ultraslow diffusion, strong
anomaly, and other complex behaviors. The overdamped motion of
a harmonic oscillator driven by distributed-order noises is investigated in
detail. A summary is presented in Section \ref{Sum}.

\section{GLE in presence of white and power-law noises}
\label{sect2}

We study the GLE for both white and power-law noises, and their combinations.
At the end of this section we also derive the overdamped limit.

\subsection{Additive white noises}
\label{Nwhite}

\begin{figure}
\resizebox{1.0\textwidth}{!}{(a)
\includegraphics{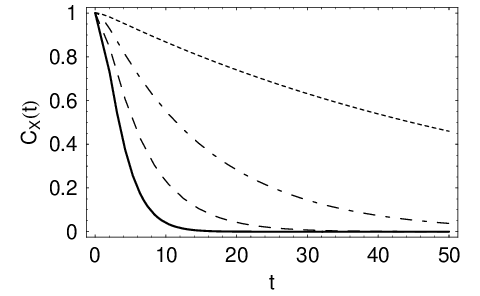} (b)
\includegraphics{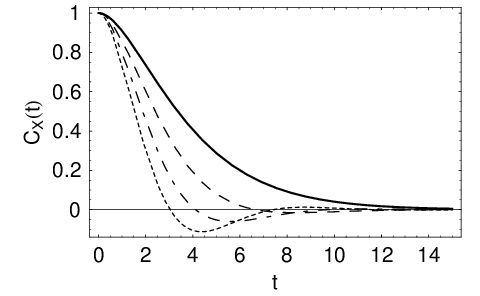}} \\ \resizebox{0.5\textwidth}{!}{(c)
\includegraphics{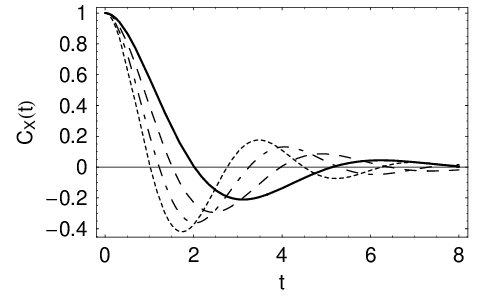}}
\caption {Normalized displacement correlation function for $\kappa=1$:
(a) overdamped motion $\kappa>2\omega$ (\ref{NDCF thermal delta}),
$\omega=\omega_{c}=1/2$ (solid line), $\omega=3/8$ (dashed line);
$\omega=1/4$ (dot-dashed line); $\omega=1/8$ (dotted line); (b)
underdamped motion (\ref{NDCF thermal delta2}),
$\omega=\omega_{c}=1/2$ (solid line), $\omega=5/8$ (dashed line);
$\omega=3/4$ (dot-dashed line); $\omega=7/8$ (dotted line); (c)
underdamped motion (\ref{NDCF thermal delta2}),
$\omega=\omega_{c}=9/8$ (solid line), $\omega=11/8$ (dashed line);
$\omega=13/8$ (dot-dashed line); $\omega=15/8$ (dotted line).}
\label{figCXdelta}
\end{figure}

The simplest case of the GLE is obtained for a test particle connected to a
thermal bath of temperature $T$ effecting $N$ additive internal white Gaussian
and zero-mean noises, equation (\ref{sum}), in which each component fulfils
$\zeta_i(t)=\delta(t)$ (i.e., $\hat{\zeta}_i(s)=1$). Physically such a joint
noise may stem from an environment with different components. As we assume
that the noise is internal, from the second FDT we conclude that the friction
memory kernel is given by
\begin{equation}
\hat{\gamma}(s)=\sum_{i=1}^N\frac{\alpha_i^2}{k_BT}
\end{equation}
in Laplace space. From relation (\ref{G(s)}), we obtain the relaxation function
$G(t)$ in the form
\begin{align}
G(t)=\mathcal{L}^{-1}\left\{\frac{1}{s^{2}+\kappa s+\omega^{2}};t\right\}
=\frac{2}{\sqrt{\kappa^2-4\omega^2}}\exp\left(-\frac{\kappa t}{2}\right)
\sinh\left(\frac{t}{2}\sqrt{\kappa^2-4\omega^2}\right),
\label{G(t) white noises h}
\end{align}
where $\kappa=\sum_{i=1}^N\alpha_i^2/(k_BT)>2\omega$.
The limit $\omega
=0$ leads to the known result $G(t)=[1-\exp(-\kappa t)]/\kappa$
\cite{sandev_tomovski_pla}, i.e., a constant diffusion coefficient in the
linear time dependence of the MSD on time in the long time limit
($t\rightarrow\infty$),
\begin{equation}
\label{D delta noise}
\langle x^2(t)\rangle\sim2\mathcal{D}t, \quad \text{where} \quad  
\mathcal{D}=\frac{\left(k_BT\right)^2}{\sum_{i=1}^N\alpha_i^2},
\end{equation}
Writing $\kappa=2\omega_c$ in terms of the compound frequency $\omega_c
=\sum_{i=1}^N\alpha_i^2/(2k_BT)$, then $\mathcal{D}(t)=t\exp(-\omega_ct)$, and
the MSD is given by
\begin{equation}
\langle x^2(t)\rangle=\omega_c^{-2}[1-(1+\omega_ct)\exp(-\omega_ct)].
\end{equation}
Therefore, the relaxation function $I(t)$ from (\ref{NDCF thermal}) with
$I(t)=\omega^{-2}\left[1-C_X(t)\right]$ is defined in terms of
\begin{equation}
\label{NDCF thermal delta}
C_X(t)=e^{-\kappa t/2}\left[\cosh\left(\frac{t}{2}\sqrt{\kappa^2-
4\omega^2}\right)+\frac{\kappa}{\sqrt{\kappa^2-4\omega^2}}\sinh\left(
\frac{t}{2}\sqrt{\kappa^2-4\omega^2}\right)\right].
\end{equation}
With our notation $\kappa=2\omega_c$ it is given by $C_X(t)=(1+\omega_ct)\exp(
-\omega_ct)$. The case $\kappa<2\omega$, i.e., the case of underdamped motion,
yields in the form
\begin{equation}
\label{NDCF thermal delta2}
C_X(t)=e^{-\kappa t/2}\left[\cos\left(\frac{t}{2}\sqrt{4\omega^2-
\kappa^2}\right)+\frac{\kappa}{\sqrt{4\omega^2-\kappa^2}}\sin\left(
\frac{t}{2}\sqrt{4\omega^2-\kappa^2}\right)\right].
\end{equation}
Thus, the MSD in the long time limit approaches the equilibrium value $\langle
x^2(t)\rangle_{\mathrm{eq}}=2k_BT/\omega^2$. A graphical representation of the
NDCF $C_X(t)$ is given in figure~\ref{figCXdelta}. From panel (a) we conclude
that in the overdamped overdamped case the NDCF shows a monotonic decay to zero,
without zero crossings. The underdamped case shows oscillatory behavior of $C_X
(t)$ with zero crossings (panels (b) and (c)), where oscillations become more
pronounced for increasing oscillator frequency.

\begin{remark}

Let us analyze the diffusion coefficient in this case somewhat further. We see
that if we consider a single internal white noise the diffusion coefficient is
$\mathcal{D}_i=(k_BT)^2/\alpha_i^2$, for $i={1,2,\dots,N}$. From relation
(\ref{D delta noise}), by using the relation between harmonic and arithmetic
mean, we obtain
\begin{equation}
\label{D delta noise limit}
\mathcal{D}=\frac{1}{\sum_{i=1}^N\frac{1}{\mathcal{D}_i}}\leq\frac{\sum_{i=1}^N
\mathcal{D}_i}{N^2}.
\end{equation}
So we see that if the particular diffusion coefficients $\mathcal{D}_i$ are
identical and equal to $\bar{\mathcal{D}}$, the diffusion coefficient for $N$
independent internal white Gaussian noise terms scales inversely to $N$,
$\mathcal{D}=\bar{\mathcal{D}}/N$.

If we write $\mathcal{D}_i=a^2/(2\tau_i)$ in a random walk-like notation, where
$a^2$ represents the squared lattice constant or the variance of the jump length
PDF \cite{stasrev,he,pccp}, we rewrite the compound diffusion coeffcient as
\begin{equation}
\mathcal{D}=\frac{1}{\sum_{i=1}^N\frac{2\tau_i}{a^2}}=\frac{a^2}{2\sum_{i=1}^N
\tau_i}=\frac{a^2}{2N\langle\tau\rangle}=\frac{\overline{\mathcal{D}}}{N}.
\end{equation}
In this sense the equality in (\ref{D delta noise limit}) always holds. We note
that the mean time $\langle\tau\rangle$ may be a misrepresentation of the largest
time scale $\tau_m\max_i\{\tau_i\}$, depending on the underlying set $\{\tau_i\}$.
\end{remark}

\subsection{Different power-law noises}\label{Npower}

We now turn to generalizing the results of \cite{sandev_tomovski_pla},
where the authors studied the GLE for free particles and $N$ independent
internal noises, by using a harmonic potential, corresonding to the Hookean
force $F(x)=-m\omega^2x$, in the GLE \eqref{GLE}. To this end we first recall
the considerations in \cite{Burov_Barkai}, where the authors study the GLE
for a harmonic potential and one internal noise, and then generalize these
results. We mention that in \cite{Burov_Barkai} complementary polynomials are
used to analyze the overdamped, underdamped, and critical behaviors of the
oscillator. Here, our investigation is based on multinomial Prabhakar-type
functions, ${\cal E}_{(\vec{\mu}),\beta}(s;\vec{a})$ (see \cite{EBazhlekova21}
and \ref{app1}). Despite the somewhat complicated notation, their
definition and use in analytic calculations is in fact quite straightforward.

\subsubsection{Power-law noises}

We first consider the single-term internal power-law noise with correlation
function $\zeta_1(t)=t^{-\lambda_1}/\Gamma(1-\lambda_1)$ (i.e., $\hat{\zeta}
_1(s)=s^{\lambda_1-1}$) for $\lambda_1\in(0,1)$. Thus, the friction memory
kernel due to the second FDT \eqref{SFDT_final} is given by $\gamma_1(t)=
\alpha_1^2\zeta_1(t)/(k_BT)$. Then the GLE \eqref{GLE} reads (using unit
mass)
\begin{equation}
\label{15/05/23-1}
\ddot{x}(t)+\frac{\alpha_1^2}{k_BT}{^C\!}D_t^{\lambda_1}x(t)+\omega^2
x(t)=\xi(t),\qquad\dot{x}(t)=v(t).
\end{equation}
Due to the power-law friction memory, the friction terms can in fact be
compactly written in terms of the fractional Caputo derivative $({^C\!}D_t^{
\lambda_1}$, and the GLE (\ref{15/05/23-1}) is often referred to as the
fractional Langevin equation \cite{Lutz,pccp}. The Caputo operator for
$\lambda_1\in(0, 1)$ is defined via ${^C\!}D_t^{\lambda_1}x(t)=\int_0^t\dot{x}
(t')(t-t')^{-\lambda_1}dt'/\Gamma(1-\lambda_1)$. The relaxation functions
$G_1(t)$, $I_1(t)$, and $g_1(t)$ are obtained by inverting the Laplace
transform \eqref{G(s)}. They are given in terms of the multinomial
Prabhakar-type functions ${\cal E}_{(\vec{\mu}),\beta}(s;\vec{a})=t^{\beta
-1}E_{(\vec{\mu}),\beta}(-a_1 t^{\mu_1},\ldots,-a_N t^\mu_N)$ where $\vec{
\mu}=\mu_1,\ldots\mu_N$ and $\vec{a}=a_1,\ldots,a_N$ (see \cite{EBazhlekova21}
and \ref{app1}) as
\begin{align}\label{22/06/23-1}
G_1(t)=\mathcal{L}^{-1}\left\{(s^2+A_1s^{\lambda_1}+\omega^2)^{-1};t\right\}
=\mathcal{E}_{(2,2-\lambda_1),2}(t;\omega^2,A_1),
\end{align}
and 
\begin{equation}
\label{16/05/23-1}
I_1(t)={\cal E}_{(2,2-\lambda_1),3}(t;\omega^2,A_1),\quad g_1(t)=\mathcal{E}_{
(2,2-\lambda_1),1}(t;\omega^2,A_1),
\end{equation}
where $A_1=\alpha_1^2/(k_BT)$. For $t\gg1$ these function are estimated by the
two-parameter ML function, see (\ref{three parameter ml}), whose asymptotic
behaviors follow from Eq.~(\ref{GML_formula}), such that
\begin{align}
\label{22/06/23-2}
G_1(t)\sim_{t\rightarrow\infty} A_1^{-1}t^{\lambda_1-1}E_{\lambda_1,\lambda_1}(-\omega^2t^{\lambda
_1}/A_1)
\sim_{t\rightarrow\infty}-\frac{A_1}{\omega^4}\frac{t^{-\lambda_1-1}}{\Gamma(-\lambda_1)}=\frac{
A_1\lambda_1}{\omega^4}\frac{t^{-\lambda_1-1}}{\Gamma(1-\lambda_1)},
\end{align}
\begin{align}
\label{16/05/23-2}
I_1(t)\sim_{t\rightarrow\infty} A_1^{-1}t^{\lambda_1}E_{\lambda_1,\lambda_1+1}(-\omega^2t^{\lambda
_1}/A_1) \sim_{t\rightarrow\infty}\frac{1}{\omega^2}\Big[1-\frac{A_1}{\omega^2}\frac{t^{-\lambda_1}}{\Gamma
(1-\lambda_1)}\Big],
\end{align}        
and
\begin{align}
\label{22/06/23-3}
g_1(t)\sim_{t\rightarrow\infty} A_1^{-1}t^{\lambda_1-2}E_{\lambda_1,\lambda_1-1}(-\omega^2
t^{\lambda_1}/A_1)
\sim_{t\rightarrow\infty}-\frac{A_1}{\omega^4}\frac{t^{-\lambda_1-2}}{\Gamma(-1-\lambda_1)}.
\end{align}
The same results can be obtained by calculating the appropriate overdamped
relaxation functions such that we have $\lim_{t\to\infty}G_1(t)=G_{o;1}(t)$,
$\lim_{t\to\infty}I_1(t)=I_{o;1}(t)$, and $\lim_{t\to\infty}g_1(t)=g_{o;1}(t)$.

\subsubsection{$N$ power-law noises}
\label{Npower2}

The compound fractional Langevin equation, i.e., the GLE in which we use $N$
independent internal noises, for $m=1$ reads
\begin{equation}
\label{16/05/23-5}
\ddot{x}(t)+\sum_{r=1}^NA_r{{^C\!}D_t^{\lambda_r}}x(t)+\omega^2x(t)=\xi(t).
\end{equation}
The friction memory kernel $\gamma_N(t)$ becomes $\sum_{r=1}^N\gamma_r(t)$,
where $\gamma_r(t)$ is defined analogously to $\gamma_1(t)$ above, but
instead of a single internal noise we now have $N$. Here, we assume that $0<
\lambda_1<\dots<\lambda_N<1$. Using short-hand notation we denote the
relaxation functions $g_N(t)$, $G_N(t)$, and $I_N(t)$ by the function $F_{N;j}
(t)$ indexed by $j=1,2,3$ such that $F_{N;1}(t)=g_N(t)$ is obtained for $j=1$,
$F_{N;2}(t)=G_N(t)$ for $j=2$, and $F_{N;3}(t)=I_N(t)$ for $j=3$. The auxiliary
functions $F_{N;j}(t)$ read
\begin{equation}
\label{16/05/23-6}
F_{N; j}(t)={\cal E}_{(2,2-\lambda_1,\ldots,2-\lambda_N),j}(t;\omega^2,A_1,
\ldots,A_N)
\end{equation}
in terms of the Prabakhar-type function. Following \cite{EBazhlekova21} we
check the behavior of $F_{N; j}(t)$ in \eqref{16/05/23-6} at long $t$. In
this case we obtain that $F_{N;j}(t)$ tends to the relaxation function for
$N=1$, i.e.,
\begin{equation}
\label{22/06/23-5}
F_{N;j}(t)\sim_{t\rightarrow\infty} A_1^{-1}t^{\lambda_1+j-3}E_{\lambda_1,\lambda_1+j-2}(-\omega
^2t^{\lambda_1}/A_1).
\end{equation}
Their further asymptotics in the long-time limit $t\to\infty$ depend on the
value of $j$, such that for $j=1,2,3$ we get Eqs.~\eqref{22/06/23-3},
\eqref{16/05/23-2}, and~\eqref{22/06/23-2}, respectively. Making use of
Eq.~\eqref{22/06/23-5} we calculate the quantities (i) to (iv) defined in
the introductory section necessary to describe the time evolution of
stochatic systems coupled to a thermal bath.  

Based on these results the MSD can be shown to be given by the expression
\begin{equation}
\label{17/05-1}
\langle x^2(t)\rangle_N=2k_BT{\cal E}_{(2,2-\lambda_1,\ldots,2-\lambda_N),3}
(t;\omega^2,A_1,\ldots,A_N).
\end{equation}
The force-free case $\omega=0$, rewritten in terms a series of three-parameter
ML functions (see \ref{app1}), yields the known result
\cite[Eqs.~(17) and~(27)]{sandev_tomovski_pla} such that we have
\begin{equation}
\label{17/05-2}
\langle x^2(t)\rangle_{N}\sim_{t\rightarrow\infty}\langle x^2(t)\rangle_{\mathrm{eq}}\left(1-\frac{
A_1}{\omega^2}\frac{t^{-\lambda_1}}{\Gamma(1-\lambda_1)}\right).
\end{equation}
The MSD approaches the equilibrium (thermal) value $\langle x^2(t)\rangle_{
\mathrm{eq}}=2k_BT/\omega^2$ in power-law fashion, instead of the exponentially
fast relaxation for the normal Orstein-Uhlenbeck process. Note that similar
power-law relaxations are known from subdiffusive continuous time random walks
\cite{Metzler1,Metzler3} and from the time-averaged MSD of the fractional
Langevin equation \cite{jae_pre,jeon1} in an external harmonic potential.
From Eq.~\eqref{17/05-2} we see that the noise with the smaller exponent
$\lambda_1$ has the dominant contribution to the oscillator behavior in the
long time limit. We also note that the same result for the MSD in the long time
limit can be obtained by Tauberian theorems (see \ref{app2}) if we analyze the
behavior of $\hat{I}_N(s)$ in the limit $s\to 0$ \cite{Gorenflo_Mainardi}. The
short time limit $t\to0$ yields $I_N(t)\simeq t^2/2-A_2t^{4-\lambda_2}/\Gamma(
5-\lambda_2)$, so that we conclude that the noise with the larger exponent
$\lambda_2$ has dominates the dynamic in the short time limit.


The DACF \eqref{NDCF thermal} in the case of thermal initial conditions $\langle
x_0^2\rangle=k_BT/\omega^2$, $\langle x_0v_0\rangle=0$, and $\langle\xi(t)x_0
\rangle=0$, from Eqs.~\eqref{NDCF thermal} and \eqref{16/05/23-6}, we obtain
\begin{equation}
\label{NDCF power law noises ex}
C_X(t)=1-\omega^2{\cal E}_{(2,2-\lambda_1,\ldots,2-\lambda_N),3}(t;\omega^2,
A_1,\ldots,A_N).
\end{equation}
Its asymptotic behavior in the long time limit via Eq. \eqref{16/05/23-2}
reads
\begin{align}
\label{NDCF power law noises ex2}
C_X(t)\sim_{t\rightarrow\infty}1-\frac{\omega^2}{A_1}t^{\lambda_1}E_{\lambda_1,\lambda_1+1}
\Big(-\frac{\omega^2}{A_1}t^{\lambda_1}\Big)=E_{\lambda_1}\Big(-\frac{\omega^2}{
A_1}t^{\lambda_1}\Big)
\sim_{t\rightarrow\infty}\frac{A_1}{\omega^2}\frac{t^{-\lambda_1}}{\Gamma(1-\lambda_1)}.
\end{align}
Thus we obtain a power-law decay, approaching the zero line from
positive values for $\lambda_1\in(0,1)$, and from negative values for
$\lambda_1\in(1,2)$, if we consider a memory kernel defined in Laplace space
as $\hat{\gamma}(s)= \sum_{i=1}^Ns^{\lambda_i-1}$, $1<\lambda_i<2$. Since
$E_{\alpha}(-x)$ is a completely monotone function for $\alpha\in(0, 1)$
\cite{oliveira,KGorska,mainardi_arxiv,Pollard}, we conclude that the
normalized displacement correlation function is completely monotone in the long
time limit for $\lambda_{1}\in(0, 1)$. A more detailed analysis of $C_X(t)$
is provided in subsection (\ref{overdamped}) for the case of high damping,
which has more practical application in the theory of anomalous dynamics
in single particle tracking and protein dynamics. 
We mention that the equality in Eq.~\eqref{NDCF power law
noises ex} was proven in \cite{KGorska18} and \cite[Remark 3]{KGorska23}.

\subsection{Combinations of white and power-law noises}
\label{white_power}
 
In the same way as for the power-law noises we can analyze the
relaxation functions for a mixture of $\delta$- and power-law distributed
noises. The approach given in \cite{sandev_tomovski_pla} can be applied
in the case of a harmonic oscillator driven by $P$ white noises and $Q$
power-law noises ($P+Q=N$) as well. Here we consider the special case
$\gamma(t)= B_1\delta(t)+B_2t^{-\lambda}/\Gamma(1-\lambda)$ where $B_1=
\alpha^2/(k_BT)$, $B_2=\beta^2/(k_BT)$, and $0<\lambda< 1$
\cite{tateishi, sandev_tomovski_pla}.

With the help of Eq.~\eqref{5/07/23-2} we see that
\begin{equation}
G(t)={\cal E}_{(2,2-\lambda,1),2}(t;\omega^2,B_2,B_1).
\end{equation}
From Eqs.~\eqref{5/07/23-3a} and \eqref{5/07/23-3b} we can calculate the
associated integral and derivative which, respectively, lead to $I(t)$ and
$g(t)$. Then, the MSD obtained from Eq.~\eqref{19/05/23-1} reads
\begin{equation}
\label{7/07/23-7}
\langle x^2(t)\rangle=2k_BT{\cal E}_{(2,2-\lambda,1),3}(t;\omega^2,B_2,B_1).
\end{equation}
Notice that Eq.~\eqref{7/07/23-7} in the force-free case $\omega=0$, after
applying Eq.~\eqref{7/07/23-6}, can be recognized as \cite[Eq.~(27) for
$\lambda_1=\lambda$, $\lambda_2=1$]{sandev_tomovski_pla}, namely,
\begin{equation}
\langle x^2(t)\rangle=2k_BT\sum_{n=0}^{\infty}\left(-B_2\right)^{n}t^{(2-
\lambda)n+2}E_{1,(2-\lambda)n+3}^{n+1}(-B_1 t).
\end{equation}
The asymptotic of the MSD at short and long times are calculated from
Eqs.~\eqref{5/07/23-5a} and \eqref{5/07/23-5b}, yielding
\begin{equation}
\label{7/07/23-8a}
\langle x^2(t)\rangle\sim_{t\rightarrow0}\frac{t^2}{2}-\frac{t^3}{3!}-\frac{t^4}{4!}-\frac{
t^{4-\lambda}}{\Gamma(5-\lambda)} 
\end{equation}
for $t\to0$, and for $t\gg1$ we have
\begin{align}
\label{7/07/23-8b}
\langle x^2(t)\rangle\sim_{t\rightarrow\infty}\frac{2k_BT}{B_2}t^{\lambda}E_{\lambda,1+\lambda}
\Big(-\frac{\omega^2}{B_2}t^{\lambda}\Big)
\sim_{t\rightarrow\infty}\frac{2k_BT}{\omega^2}\left[1-\frac{B_2}{\omega^2}\frac{t^{-\lambda}}{
\Gamma(1-\lambda)}\right],
\end{align}
Which means that the MSD has a power-law decay to the equilibrium value $\langle
x^2(t)\rangle_{\mathrm{eq}}=2k_BT/\omega^2$. We conclude that the power-law noise
is dominant in the long time limit, and the white noise in the short time limit,
as naively expected.

For the DACF $C_X(t)$, we obtain from relation \eqref{NDCF thermal} that
\begin{equation}
\label{NDCF delta noise power law noise ex} 
C_X(t)=1-\omega^2{\cal E}_{(2,2-\lambda,1),3}(t;\omega^2,B_2,B_1), 
\end{equation} 
from where the long time limit follows,
\begin{align}
\label{NDCF delta noise power law noise ex long time} 
C_X(t)\sim_{t\rightarrow\infty}1-\frac{\omega^2}{B_2}t^{\lambda}E_{\lambda,1+\lambda}\Big(
-\frac{\omega^2}{B_2}t^{\lambda}\Big)\sim_{t\rightarrow\infty}\frac{B_2}{\omega^2}\frac{t^{-\lambda}}{\Gamma(1-\lambda)}.
\end{align} 
We conclude that $C_X(t)$ in the long time limit is a completely monotone
function since $0<\lambda<1$, showing a power-law decay to zero.

Following the same procedure one may consider mixtures of white noises,
power-law noises, and ML type noises. The calculation of the relaxation
function can be represented in terms of multinomial Prabhakar functions
\cite{sandev_tomovski_pla, EBazhlekova21}.

\subsection{Overdamped limit}\label{overdamped}

Next, we analyze the high-damping limit, in which the inertial term $m\ddot{x}
(t)$ can be neglected. Then, the relaxation functions are defined by
Eqs.~\eqref{12/06/23-1}.

The case of a mixture of $N$ internal white noises considered in \ref{Nwhite},
for the MSD $\langle x^2(t)\rangle_o=2k_BTI_o(t)$ and the DACF $C_{X,o}(t)$
yields
\begin{align}
\label{9/07/23-1}
\langle x^2(t)\rangle_o &=2k_BT\mathcal{L}^{-1}\left\{\frac{s^{-1}}{\kappa s
+\omega^2};t\right\}=\frac{2k_BT}{\omega^2}\left[1-\exp\left(-\frac{\omega^2}{
\kappa}t\right)\right],\\
\label{9/07/23-2}
C_{X,o}(t)&=\exp\Big(-\frac{\omega^2}{\kappa}t\Big)=\exp\left(-\frac{\mathcal{D}
\omega^2}{k_BT}t\right),
\end{align}
where $\kappa$ is given below Eq.~\eqref{G(t) white noises h} and ${\cal D}$ is
defined in Eq.~\eqref{D delta noise}. We find that $C_{X,o}(t)$ has a monotonic
exponential decay, as expected.

For the case of $N$ power-law noises, it follows from relation \eqref{19/05/23-1}
in the overdamped case that
\begin{align}
\label{I(t) overdamped power law noises}
\langle x^2(t)\rangle_o&=\frac{2k_BT}{A_N}{\cal E}_{(\lambda_N,\lambda_N-\lambda
_{1},\ldots,\lambda_N-\lambda_{N-1});\lambda_N+1}\Big(t,\frac{\omega^2}{A_N},
\frac{A_{1}}{A_N},\ldots,\frac{A_{N-1}}{A_N}\Big),
\end{align}
\begin{align}\label{ndcf n noises}
C_{X,o}(t)&=1-\frac{\omega^2}{A_N}{\cal E}_{(\lambda_N,\lambda_N-\lambda_{1},
\ldots,\lambda_N-\lambda_{N-1});\lambda_N+1}\Big(t,\frac{\omega^2}{A_N},
\frac{A_{1}}{A_N},\ldots,\frac{A_{N-1}}{A_N}\Big).
\end{align}
The DACF shows an asymptotic power-law decay in the long time limit, i.e.,
\begin{equation}
\label{CX(t) overdamped power law noises long}
C_{X,o}(t)\sim_{t\rightarrow\infty} E_{\lambda_1}\left(-\frac{\omega^2}{A_1}t^{\lambda_1}\right)
\sim_{t\rightarrow\infty}\frac{A_{1}}{\omega^2}\frac{t^{-\lambda_1}}{\Gamma\left(1-\lambda_1
\right)}.
\end{equation}
Thus, we can use equation (\ref{GLE overdamped}) in the overdamped limit case,
which is simpler instead of the GLE (\ref{GLE}), to analyze the asymptotic
behavior of the harmonic oscillator in the long time limit.

\begin{figure}
\resizebox{1.0\textwidth}{!}{(a)
\includegraphics{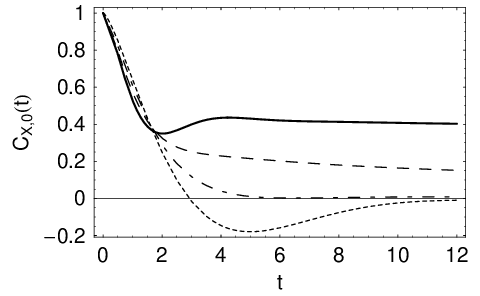} (b)
\includegraphics{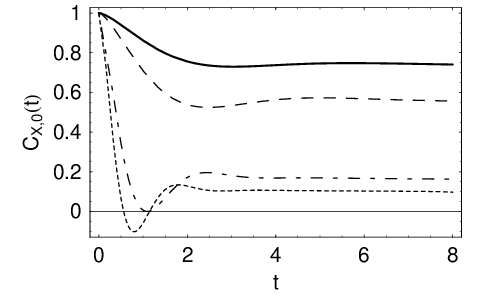}} \\ \resizebox{0.5\textwidth}{!}{(c)
\includegraphics{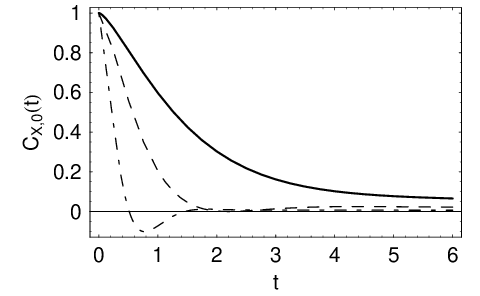}}
\caption{Normalized displacement correlation function (\ref{ndcf n noises}) for the following cases for $N=2$:
(a) $\omega=1$, $\lambda_{2}=3/2$, $\lambda_{1}=1/8$, (solid line),
$\lambda_{1}=1/2$ (dashed line), $\lambda_{1}=7/8$ (dot-dashed
line), $\lambda_{1}=5/4$ (dotted line); (b) $\lambda_{1}=1/8$,
$\lambda_{2}=3/4$, $\omega=0.5$ (solid line), $\omega=0.75$ (dashed
line), $\omega=1.8959706$ (dot-dashed line), $\omega=2.5$ (dotted
line); (c) $\lambda_{1}=3/4$, $\lambda_{2}=3/2$, $\omega=1$ (solid
line), $\omega=1.617$ (dashed line), $\omega=3$ (dot-dashed line). }
\label{fig_powerlaw}
\end{figure}

A graphical representation of the DACF is presented in figure~\ref{fig_powerlaw}.
From panel (a) we see that by changing the values of the parameters $\lambda_1$
and $\lambda_2$ for fixed frequency $\omega$ there appears a non-monotonic
decay of $C_{X,o}(t)$ without zero crossings, approaching the zero line
at infinity (see solid line), a monotonic decay without zero crossings to
zero at infinity (dashed line), a non-monotonic decay without zero crossings
approaching zero at a finite time instant and at infinity (dot-dashed line),
or a non-monotonic decay with zero crossings approaching zero at infinity
(dotted line). All long time decays of the DACF are of power-law form to
zero, as it can be anticipated from relation (\ref{CX(t) overdamped power
law noises long}). The behavior of $C_X(t)$ for different values of the
frequency $\omega$ and fixed values of $\lambda_1$ and $\lambda_2$ is shown in
figure~\ref{fig_powerlaw}(b) and (c).  We see that there are different critical
frequencies, i.e., the frequency at which $C_{X,o}(t)$ changes its behavior,
for instance, from non-monotonic to monotonic decay without zero crossings,
or the frequency at which $C_{X,o}(t)$ crosses the zero line. Such different
types of critical frequencies were discussed in \cite{Burov_Barkai}.

\begin{figure}
\resizebox{1.0\textwidth}{!}{(a)
\includegraphics{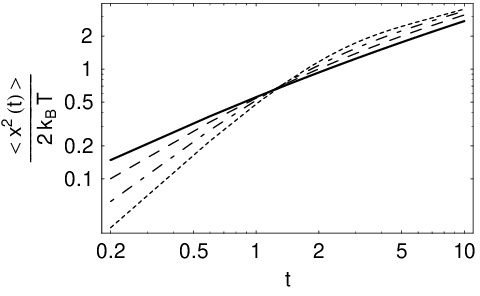} (b)
\includegraphics{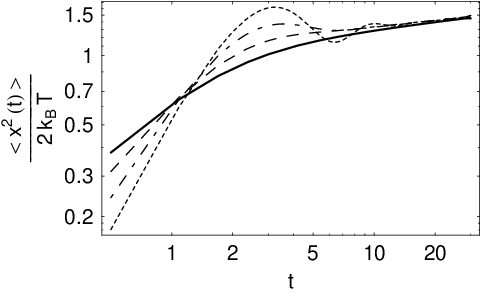}} \\ \resizebox{0.5\textwidth}{!}{(c)
\includegraphics{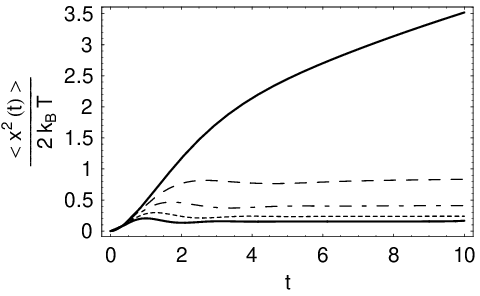}}
\caption{MSD (\ref{I(t) overdamped power law noises}) for the following cases for $N=2$:
(a) $\omega=0$, $\lambda_{1}=1/2$, $\lambda_{2}=1$, (solid line),
$\lambda_{2}=5/4$ (dashed line), $\lambda_{2}=3/2$ (dot-dashed
line), $\lambda_{2}=7/4$ (dotted line); (b) $\omega=0$,
$\lambda_{1}=0.1$, $\lambda_{2}=1$, (solid line),
$\lambda_{2}=5/4$ (dashed line), $\lambda_{2}=3/2$ (dot-dashed
line), $\lambda_{2}=7/4$ (dotted line); (c) $\lambda_{1}=1/2$,
$\lambda_{2}=7/4$, $\omega=0$ (upper solid line), $\omega=1$
(dashed line), $\omega=1.5$ (dot-dashed line), $\omega=2$ (dotted
line), $\omega=2.5$ (lower solid line).}
\label{fig_MSD_omega_omega0_powerlaw}
\end{figure}

Figure~\ref{fig_MSD_omega_omega0_powerlaw} depicts the MSD for unconfined and
confined motion. For free motion the MSD may have monotonic, non-monotonic,
and oscillatory behavior, turning into a power-law form in the long time
limit. For the confined case the MSD has different behaviors at intermediate
times, and in the long time limit the MSD has a power-law approach to the
equilibrium value $\langle x^2(t)\rangle_{\mathrm{eq}}=2k_BT/\omega^2$.

In the same way as above we obtain the following DACF for the case of a mixture
of a white noise a and power-law noise (see results \ref{white_power}),
\begin{eqnarray}
\label{CX(t) overdamped white power law noises1}
C_{X,o}(t)=\sum_{n=0}^{\infty}\left(-\frac{\omega^2}{B_1}\right)^nt^n
E_{1-\lambda,n+1}^n\left(-\frac{B_2}{B_1}t^{1-\lambda}\right),
\end{eqnarray}
for $0<\lambda<1$, and
\begin{eqnarray}
\label{CX(t) overdamped white power law noises2}
C_{X,o}(t)=\sum_{n=0}^{\infty}\left(-\frac{\omega^2}{B_2}\right)^nt^{\lambda
n}E_{\lambda-1,\lambda n+1}^n\left(-\frac{B_1}{B_2}t^{\lambda-1}\right),
\end{eqnarray}
for $1<\lambda<2$, where $B_1$ and $B_2$ are given at the beginning of Sec.
\ref{white_power}. These results are equivalent to those obtained in subsection
\ref{white_power} in the long time limit for the GLE when the inertial term is
not neglected, as it should be.

\section{Distributed-order Langevin equations}
\label{sect3}

It has been shown that the distributed-order differential equations
are suitable tool for modeling ultraslow relaxation and diffusion
processes
\cite{chechkin1,chechkin2,Eab2,kochubei,mainardi_book,Mainardi_distributed,sandev_tomovski_pla}.
In case of distributed order differential equations one uses the
following memory kernel (see for example
\cite{Mainardi_distributed})
\begin{equation}\label{distributed noise p}
\gamma(t) = (k_{B}T)^{-1}\, \int_{0}^{1}p(\lambda)\frac{t^{-\lambda}}{\Gamma(1-\lambda)}d\lambda,
\end{equation}
where $p(\lambda)$ is a dimensionless, non-negative weight function with
$\int_0^1p(\lambda)\mathrm{d}\lambda=c$, where $c$ is a constant. When $c=1$,
$p(\lambda)$ is normalized. If we substitute the distributed-order memory kernel
in the GLE (\ref{GLE}) we transform it to the following distributed-order
Langevin equation
\begin{eqnarray}
\label{GLE distributed order}
\ddot{x}(t)+\frac{1}{k_BT}\int_0^1p(\lambda)^{C\!}D^{\lambda}_t x(t)d\lambda+\frac{dV(x)}{dx}=\xi(t),\quad\dot{x}(t)=v(t).
\end{eqnarray}
Here we note that assumption (\ref{assumption}) is satisfied for
distributed-order Langevin equations since $\lim_{s\rightarrow0}s\hat{\gamma}(s)
\simeq\lim_{s\rightarrow0}\int_0^1p(\lambda)s^{\lambda}d\lambda=0$.
Thus we can use the representations of MSD, VACF, time-dependent diffusion
coefficient and DACFs in terms of the relaxation functions.

We now consider the distributed-order Langevin equation (\ref{GLE distributed
order}) in the presence of the constant external force $F$. Then we have
\begin{eqnarray}
\label{GLE constantF}
\ddot{x}(t)+\frac{1}{k_BT}\int_0^1p(\lambda)^{C\!}D^{\lambda}_tx(t)\,d\lambda-F
=\xi(t),\quad\dot{x}(t)=v(t),
\end{eqnarray}
from which, by the Laplace transform method, we obtain
\begin{equation}
\label{x F}
x(t)=\langle x(t)\rangle_F+\int_0^tG(t-t')\xi(t')dt',
\end{equation}
where $\langle x(t)\rangle_F=FI(t)$ and $\hat{\gamma}(s)$ is given by
Eq.~\eqref{distributed noise p},
and where
\begin{equation}
\label{7/07/23-5}
G(t)=\mathcal{L}^{-1}\left\{\frac{1}{s^2+s\hat{\gamma}(s)};t\right\}\quad
\text{and}\quad I(t)=\mathcal{L}^{-1}\left\{\frac{s^{-1}}{s^{2}+s\hat{\gamma}
(s)};t\right\}.
\end{equation}
The latter expression corresponds to (\ref{G(s)}) for a free particle ($\omega=0$).
Thus, we conclude that the generalized Einstein relation $\langle x^2(t)\rangle_F=
[F/(2k_BT)]\langle x^2(t)\rangle_{F=0}$ is satisfied for the distributed-order
Langevin equation (\ref{GLE distributed order}), where $\langle x^2(t)\rangle_{F=0}
=2k_BT\mathcal{L}^{-1}\{s^{-1}[s^2+s\hat{\gamma}(s)]^{-1};t\}$ is the MSD for the
case of a free particle.

\subsection{Force-free case}

We first consider the distributed-order Langevin equation for a free particle with
$\omega=0$. Note that a weight function of the form $p(\lambda)=\sum_{i=1}^N\alpha
_i^2\delta(\lambda-\lambda_i)$ yields the fractional Langevin equation and we have the compound white Gaussian noise, see Subsect.~\ref{Npower2}. For $p(\lambda)=\alpha^2$ we obtain the uniformly distributed noise \cite{kochubei}
\begin{equation}
\label{distributed noise p uniform}
k_BT\gamma(t)=\alpha^2\int_0^1\frac{t^{-\lambda}}{\Gamma\left(1-\lambda\right)}
d\lambda,
\end{equation}
which was used by Kochubei in the theory of evolution equations. In
\cite{sandev_tomovski_pla} we analyzed the GLE for a free particle with
friction memory kernel of distributed order (\ref{distributed noise
p uniform}) and showed that the MSD in the long time limit is given by
\begin{equation}
\label{MSD power law noise distributed asymptotic}
\langle x^2(t)\rangle\sim_{t\to\infty}\frac{2\left(k_BT\right)^2}{\alpha^2}\Big[C+\log t
+{\rm e}^t\mathrm{E}_1(t)\Big],
\end{equation}
where $C = 0.577216$ is the Euler-Mascheroni (or Euler's) constant, $\mathrm{Ei}
(-t)=-\int_t^{\infty}(e^{-x}/x)dx$ is the exponential integral
\cite{erdelyi}, and $\mathrm{E}_1(t)=-\mathrm{Ei}(-t)$. From the asymptotic
expansion ${\rm E}_1(t)\sim_{t\to\infty} t^{-1}{\rm e}^{-t}\sum_{k=0}^{n-1}(-1)^kk!t^{-k}$
\cite{mainardi_book}, which has an error of order ${\cal O}(n!t^{-n})$, we
obtained that the particle shows ultraslow diffusion, i.e., $\langle x^2(t)\rangle\sim_{t\to\infty}
[2(k_BT)^2/\alpha^2](C+\log t)$ \cite{sandev_tomovski_pla}.

Consider now the power-law case $p(\lambda)=\alpha^2\beta\lambda^{\beta-1}$,
where $\beta > 0$. Then the friction memory kernel becomes
\begin{equation}
\label{distributed noise p nu}
k_BT\gamma(t)=\alpha^2\int_0^1\beta\lambda^{\beta-1}\frac{t^{-\lambda}}{\Gamma(
1-\lambda)}d\lambda.
\end{equation}
For the MSD in the long time limit, we obtain (see (\ref{7/07/23-5}) based on
Tauberian theorems)
\begin{equation}
\label{msd distributed nu long time final}
\langle x^2(t)\rangle\sim_{t\to\infty}\frac{2(k_BT)^2}{\alpha^2}\frac{\log^{\beta}t}{
\Gamma(1+\beta)}.
\end{equation}
From this result we conclude that this power-law model leads to ultraslow diffusion (for $\beta=1$)
or a strong anomaly (in Ref.~\cite{drager} a strong anomaly means the behavior of form $\langle x^2(t)\rangle\sim\log^{\nu}t$, which for $\nu=4$ has the same form as the MSD of the Sinai
diffusion model in a random force field \cite{aljaz,sinai}, compare also to the
discussion in \cite{igorg}). This result for the MSD is equivalent to the one
obtained by Eab and Lim \cite{Eab2}, since the long time limit considered here
corresponds to the case of the overdamped limit (the inertial term $\ddot{x}(t)$
is neglected) studied in \cite{Eab2}. The same result for the MSD can be obtained
from the distributed-order diffusion equation \cite{chechkin2} in which the weight
function corresponds to the one considered in the memory kernel
(\ref{distributed noise p nu}).

Next we consider the weight function $p(\lambda)=\alpha^2/(\lambda_2-\lambda_1)$
for $0\leq\lambda_1<\lambda<\lambda_2\leq1$, and $p(\lambda)=0$ otherwise
\cite{chechkin1}. With the help of relation \eqref{7/07/23-5} the MSD yields in
the form
\begin{align}
\label{msd distributed lambda21 long time final}
\langle x^2(t)\rangle=2k_BT\left[\frac{t^2}{2}+\sum_{n=1}^{\infty}\Big(-\frac{
\alpha^2}{k_BT}\Big)^n\frac{1}{(\lambda_2-\lambda_1)^n}\sum_{k=0}^n\binom{n}{k}
(-1)^k \mu(t,n-1,(2-\lambda_2)n+(\lambda_2-\lambda_1)k+2)\right],
\end{align}
which in the long time limit becomes (use the Tauberian theorem in
Eq.~(\ref{7/07/23-5}))
\begin{align}
\langle x^2(t)\rangle\sim_{t\rightarrow\infty}\frac{2\left(k_BT\right)^2}{\alpha^2}\sum_{n=1}^{
\infty}\frac{t^{\lambda_2-(\lambda_2-\lambda_1)n}}{\Gamma(\lambda_2-(\lambda_2-
\lambda_1)n+1)}\left[\ln t^{\lambda_2-\lambda_1}-(\lambda_2-\lambda_1)\psi(\lambda_2-
(\lambda_2-\lambda_1)n+1)\right],
\end{align}
where we used the Volterra function $\mu\left(t,\beta,\alpha\right)$ defined in~\ref{app3}, and $\psi=\frac{\Gamma'}{\Gamma}$ is the digamma function. Thus, we conclude
that in the long time limit the MSD has the ultraslow form
\begin{equation}
\langle x^2(t)\rangle\sim_{t\rightarrow\infty}\frac{2(k_BT)^2}{\alpha^2}\frac{\lambda_2-\lambda_1}
{\Gamma(1+\lambda_1)}t^{\lambda_1}\ln t.
\end{equation}
For $\lambda_1=0$ and $\lambda_2=1$ we arrive at result (\ref{distributed noise
p uniform}) obtained for uniformly distributed noise with a logarithmic MSD
dependence on time.

Such relaxation patterns of logarithmic form and ultraslow diffusion have been
observed in the analysis of distributed-order relaxation and diffusion equations
by Kochubei \cite{kochubei2,kochubei3,kochubei4,kochubei} as well as by
Mainardi et al. \cite{mainardi_book,Mainardi_distributed}.

\subsection{Harmonic oscillator: overdamped limit}

We now move on to consider the distributed-order GLE for an harmonic oscillator
with different weight functions in the overdamped limit. For the uniformly
distributed noise (\ref{distributed noise p uniform}) we obtain the following
form of the MSD
\begin{align}
\label{MSD overdamped uniformly distributed}
\langle x^2(t)\rangle_o&=2k_BT\mathcal{L}^{-1}\left\{\frac{s^{-1}}{\frac{
\alpha^2}{k_BT}\frac{s-1}{\log s}+\omega^{2}};t\right\}=2k_BT\mathcal{L}^{-1}\left\{\frac{s^{-1}\log s}{\omega^{2}\log s+As-A};t
\right\},
\end{align}
where $A=\alpha^2/(k_BT)$. From Tauberian theorems (see ~\ref{app2}) we
analyze the asymptotic behavior in the long and short time limits. At long
times ($t\rightarrow\infty$, equivalent to $s\rightarrow0$) it follows that
\begin{align}
\label{MSD overdamped uniformly distributed longtime}
\langle x^2(t)\rangle_o\sim_{t\rightarrow\infty}\frac{2k_BT}{\omega^2}\mathcal{L}^{-1}\left\{
\frac{s^{-1}}{1-\frac{A}{\omega^2\log s}};t\right\}\sim_{t\rightarrow\infty}\frac{2k_BT}{\omega^2}
\mathcal{L}^{-1}\left\{\frac{1}{s}+\frac{\frac{A}{\omega^{2}}}{s\log s}; t
\right\}
=\frac{2k_BT}{\omega^2}\left[1+\frac{A}{\omega^2}\nu(t)\right],
\end{align}
where $\nu(t)$ is the Volterra function, see \ref{app3}.
Thus, the DACF in the long time limit becomes
\begin{eqnarray}
\label{CX overdamped uniformly distributed longtime}
C_{X,o}(t)\sim_{t\rightarrow\infty}-\frac{A}{\omega^2}\nu(t).
\end{eqnarray}
At short times ($t\to0$, or $s\to\infty$), we find for the MSD and DACF that
\begin{equation}
\langle x^2(t)\rangle_o\sim_{t\rightarrow0}\frac{2k_BT}{A}\mathcal{L}^{-1}\left\{\frac{\log
s}{s^2};t\right\}=\frac{2k_BT}{A}t\left(\log\frac{1}{t}+1-\gamma\right),\quad
\end{equation}
and
\begin{equation}
\label{CX overdamped uniformly distributed shorttime}
C_{X,o}(t)\sim_{t\rightarrow0}1-\frac{\omega^2}{A}t\left(\log\frac{1}{t}+1-\gamma\right),
\end{equation}
respectively.

Now consider the distributed-order noise (\ref{distributed noise p nu}). From
the relaxation function $I_{o}(t)$ the MSD becomes
\begin{equation}
\label{MSD overdamped lnu distributed}
\langle x^2(t)\rangle_o=\frac{2k_BT}{\omega^2}\mathcal{L}^{-1}\left\{\frac{s^{
-1}}{\frac{A}{\omega^2}\nu(\log\frac{1}{s})^{-\nu}\gamma(\nu,-\log s)+1};t\right\},
\end{equation}
where $\gamma(a,\sigma)=\int_0^\sigma t^{a-1}\exp(-t) dt$ is the lower incomplete
Gamma function. Taking the long time limit yields
\begin{align}
\langle x^2(t)\rangle_o\sim_{t\rightarrow\infty}\frac{2k_BT}{\omega^2}\left\{1-\frac{A}{
\omega^2}\Gamma(1+\nu)\mathcal{L}^{-1}\left\{\frac{1}{s}{\left(\log\frac{1}{s}
\right)^{-\nu}}; t\right]\right\}\sim_{t\rightarrow\infty}\frac{2k_BT}{\omega^2}\left[1-\frac{A}{\omega^2}\frac{\Gamma(1+\nu)}{
\log^{\nu}t}; t\right],
\label{MSD overdamped nu distributed longtime}
\end{align}
from which we obtain the asymptotic behavior of the DACF,
\begin{equation}
\label{CX overdamped nu distributed longtime}
C_{X,o}(t)\sim_{t\rightarrow\infty}\frac{2k_BT}{\omega^2}\frac{A}{\omega^2}\frac{\Gamma(1+\nu)}{
\log^{\nu}t}.
\end{equation}
In the short time limit we find
\begin{equation}
\label{MSD overdamped nu distributed shorttime}
\langle x^2(t)\rangle_o\sim_{t\rightarrow0}\frac{2k_BT}{A}\frac{\log^{\nu}\frac{1}{t}}{\Gamma(
1+\nu)}
\end{equation}
and
\begin{equation}
C_{X,o}(t)\sim_{t\rightarrow0}1-\frac{\omega^2}{A}\frac{\log^{\nu}\frac{1}{t}}{\Gamma(1+\nu)}.
\end{equation}

In a similar way we obtain for the distributed-order noise \eqref{distributed
noise p} with weight function $p(\lambda)=\alpha^2/(\lambda_2-\lambda_1)$ ($0
\leq\lambda_1<\lambda<\lambda_2\leq1$, and $p(\lambda)=0$ otherwise) the
asymptotic forms of the MSD:
\begin{equation}
\label{MSD overdamped lambda21 distributed longtime final}
\langle x^2(t)\rangle_o\sim_{t\rightarrow\infty}\frac{2k_BT}{\omega^2}\left\{1-\frac{A/\omega^2}{
\left(\lambda_2-\lambda_1\right)}\big[\nu(t,-\lambda_2)-\nu(t,-\lambda_1)\big]
\right\},
\end{equation}
in the long time limit, and
\begin{align}
\label{MSD overdamped lambda21 distributed shortime final}
\langle x^2(t)\rangle_o\sim_{t\rightarrow0}\frac{2k_BT}{A}\sum_{n=0}^{\infty}\frac{t^{\left(
\lambda_2-\lambda_1\right)n+\lambda_2}}{\Gamma[(\lambda_2-\lambda_1)n+\lambda_2
+1]}\left[\log t^{-\left(\lambda_2-\lambda_1\right)}-\left(\lambda_2-\lambda
_1\right)\psi\left(\left(\lambda_2-\lambda_1\right)n+\lambda_2+1\right)\right]
\end{align}
in the short time limit. From here we easily can find the DACF.

From these results we conclude that the distributed-order GLE may be used to
model various anomalous diffusive behaviors, such as ultraslow diffusion,
strong anomaly, and other complex diffusive regimes.

\section{Further generalizations}

\subsection{LE with power-logarithmic distributed order noises}

Let us consider the Langevin equation \eqref{GLE} with the logarithmically
distributed-order friction kernel
\begin{equation}
\label{12/07/23-1}
\gamma(t)=(k_BT)^{-1}\int_0^1\Gamma(3/2-\lambda)\frac{\log^{\lambda-1}t}{\sqrt
t}d\lambda\quad\mbox{with } \left(\hat{\gamma}(s)=\frac{\pi}{k_BT}\frac{
s-1}{\sqrt{s}\log s}\right),
\end{equation}
which satisfies the condition \eqref{assumption}. The relaxation function $I(t)$
defined by Eq. \eqref{G(s)} assumes the form
\begin{equation}
\label{13/07/23-10}
I(t)=\mathcal{L}^{-1}\left\{\frac{s^{-1}}{s^2+\frac{\pi}{k_BT}\frac{s(s-1)}{
\sqrt{s}\log{s}}+\omega^2};t\right\}. 
\end{equation}
As it is challenging to calculate the exact form of Eq.~\eqref{13/07/23-10} for
general $t>0$ we concentrate on the asymptotic behavior for small $s\ll1$, yielding
\begin{align*}
\hat{I}(s)&\sim_{s\rightarrow0}\frac{s^{-1}}{\omega^2}\frac{1}{1-\frac{\pi}{k_BT\omega^2}\frac{
s(1-s)}{\sqrt{s}\log s}}
\simeq\frac{1}{\omega^2}\left[1+\frac{\pi}{k_BT\omega^2}\frac{1}{\sqrt{s}\log s}
-\frac{\pi}{k_BT\omega^2}\frac{\sqrt{s}}{\log s}\right],
\end{align*}
where we take the zeroth and first-order terms of the series expansion. If $s$
tends zero then the term $\sqrt{s}/\log s$ can be neglected. Thus, we have
\begin{align}
\label{31/07/23-1}
\hat{I}(s)\sim_{s\rightarrow0}\frac{1}{\omega^2}\left[1+\frac{\pi}{k_BT\omega^2}\frac{1}{\sqrt{
s}\log s}\right] \quad \text{ and } \quad I(t)\sim_{t\rightarrow\infty}\frac{1}{\omega^2}\left[1+\frac{\pi}{k_BT
\omega^2}\nu(t,-1/2)\right].
\end{align}
This allows us to find the MSD and DACF for $t\rightarrow\infty$,
\begin{equation*}
\langle x^2(t)\rangle\sim_{t\rightarrow\infty}\frac{2k_BT}{\omega^2}+\frac{2\pi}{\omega^4}\nu(t,
-1/2) \quad \text{ and } \quad C_X(t)\sim_{t\rightarrow\infty}-\frac{\pi}{k_BT\omega^2}\nu(t,-1/2).
\end{equation*}

Note that for $\omega=0$ we get 
\begin{align}
\label{1/08/23-1}
\langle x^2(t)\rangle&=2k_BT\mathcal{L}^{-1}\left\{\frac{s^{-1}}{s^2+\frac{\pi}{
k_BT}\frac{s(s-1)}{\sqrt s\log s}};t\right\}\sim_{t\rightarrow\infty}2k_BT\mathcal{L}^{-1}\left\{
\frac{s^{-1}}{\frac{\pi}{k_BT}\frac{s(s-1)}{\sqrt s\log s}};t\right\}\nonumber\\
&\sim_{t\rightarrow\infty}\frac{2(k_BT)^2}{\pi}\mathcal{L}^{-1}\left\{\frac{\log s}{s^{3/2}(s-1)};t
\right\}\sim_{t\rightarrow\infty}\frac{2(k_BT)^2}{\pi}\mathcal{L}^{-1}\left\{\frac{\log \frac{1}{s}}
{s^{3/2}};t\right\}\nonumber\\
&\sim_{t\rightarrow\infty}\frac{2(k_BT)^2}{\pi}\frac{1}{\Gamma(3/2)}\sqrt{t}\log t=\frac{4(k_BT)^2}
{\pi^{3/2}}\sqrt{t}\log t.
\end{align}

\subsection{Langevin equation with distributed-order Mittag-Leffler noise}

In the next example, we study the GLE for an harmonic oscillator with the ML
friction memory kernel
\begin{equation}
\gamma(t)=\frac{1}{k_BT}\int_0^1E_{\lambda}\left(-t^{\lambda}\right)d\lambda\quad\left(\hat{\gamma}(s)=\frac{1}{k_BT}\frac{\log\frac{s+1}{2}}{s\log
s}\right),
\end{equation}
which satisfies the condition \eqref{assumption}. The long time asymptotic of
the relaxation function $I(t)$ involved in the MSD and the DACF can be calculated
by use of the Tauberian theorem (~\ref{app2}) in which we consider the limit
$s\to0$ of $\hat{I}(s)$. Thus, we have
\begin{align}
\label{1/08/23-6}
I(t)=\mathcal{L}^{-1}\left\{\frac{s^{-1}}{s^2+\frac{1}{k_BT}\frac{\log\frac{s+
1}{2}}{\log s}+\omega^2};t\right\}\sim_{t\rightarrow\infty}\frac{1}{\omega^2}\mathcal{L}^{-1}
\left\{\frac{1}{s}-\frac{1}{k_BT\omega^2}\frac{\log2}{s\log s};t\right\}=\frac{1}{\omega^2}\left[1-\frac{\log2}{k_BT\omega^2}\nu(t)\right].
\end{align}
In the force-free case with $\omega=0$ we observe ultraslow diffusion, since the
relaxation function $I(t)$ has a logarithmic time dependence,
\begin{equation}
\label{1/08/23-7}
I(t)\sim_{t\rightarrow\infty}\frac{k_BT}{\log 2}\mathcal{L}^{-1}\left\{\frac{\log\frac{1}{s}}{s};
t\right\}=\frac{{k_BT}}{\log 2}\log t=k_BT\log_2 t.
\end{equation}

\subsection{Distributed-order Langevin equation with Caputo-Prabhakar derivative}

In this part we now consider the GLE \eqref{GLE} for an harmonic oscillator with
the choice of the distributed-order Prabhakar friction memory kernel
\begin{equation}
\label{2/08/23-5}
\gamma(t)=\frac{1}{k_BT}\int_0^1t^{-\lambda}E_{{\rho},1-\lambda}^{\delta}(-t^{
{\rho}})d\lambda\quad \text{ with }\quad \hat{\gamma}(s)=\frac{1}{k_BT}\frac{s-1}{
s\log s}(1+{s^{-\rho}})^{-\delta}.
\end{equation}
We consider two cases, the first one for $0<\rho,\delta<1$ and the second one
for $-1<\rho,\delta <0$. In both cases the condition \eqref{assumption} is
satisfied. Using the Tauberian theorem we calculate the asymptotics of the
relaxation function $I(t)$. 

\noindent
{\bf (i)} In the case $0<\rho,\delta<1$ and for $t\to\infty$ we have
\begin{align}
\label{2/08/23-1}
I(t)&\sim_{t\rightarrow\infty}\mathcal{L}^{-1}\left\{\frac{s^{-1}}{\omega^2+\frac{1}{k_BT}\frac{
s-1}{\log s}(1+s^{-\rho})^{-\delta}};t\right\}\sim_{t\rightarrow\infty}\mathcal{L}^{-1}\left\{
\frac{s^{-1}}{\omega^2+\frac{1}{k_BT}\frac{s-1}{\log s}s^{\rho\delta}};t\right\}
\nonumber\\
&\sim_{t\rightarrow\infty}\frac{1}{\omega^2}\mathcal{L}^{-1}\left\{\frac{s^{-1}}{1+\frac{1}{k_BT
\omega^2}\frac{s^{\rho\delta}}{\log\frac{1}{s}}};t\right\}\sim_{t\rightarrow\infty}\frac{1}{\omega
^2}\mathcal{L}^{-1}\left\{\frac{1}{s}-\frac{1}{k_BT\omega^2}\frac{s^{-1+\rho
\delta}}{\log\frac{1}{s}};t\right\}\nonumber\\
&\sim_{t\rightarrow\infty}\frac{1}{\omega^2}\left[1-\frac{1}{k_BT\omega^2\Gamma(1-\rho\delta)}
\frac{t^{-\rho\delta}}{\log{t}}\right].
\end{align}
From the asymptotic form of $I(t)$ we obtain the long time behaviors of the MSD
and DACF,
\begin{equation}
\langle x^2(t)\rangle\sim_{t\rightarrow\infty}\frac{2k_BT}{\omega^2}-\frac{1}{\omega^4\Gamma(1-
\rho\delta)}\frac{t^{-\rho\delta}}{\log t}
\end{equation}
and
\begin{equation}
C_X(t)\sim_{t\rightarrow\infty}\frac{1}{k_BT\omega^2\Gamma(1-\rho\delta)}\frac{t^{-\rho\delta}}{
\log{t}}.
\end{equation}

In addition, note that in the free particle case ($\omega=0$) the long time limit
(or for the overdamped case when we neglect the term with $s^2$) becomes
\begin{align}
I(t)&\sim_{t\rightarrow\infty} k_BT\mathcal{L}^{-1}\left\{\frac{s^{-1}(1+{s^{-\rho}})^{\delta}\log
s}{{s-1}};t\right\}\sim_{t\rightarrow\infty} k_BT\mathcal{L}^{-1}\left\{\frac{\log s}{s^{1+\rho
\delta}(s-1)};t\right\}\nonumber\\
&\sim_{t\rightarrow\infty} k_BT\mathcal{L}^{-1}\left\{s^{-1-\rho\delta}\log\frac{1}{s};t\right\}
\sim_{t\rightarrow\infty} k_BT\frac{t^{\rho\delta}}{\Gamma(1+\rho\delta)}\log{t}.
\end{align}

\noindent
{\bf (ii)} Replacing $\rho$ by $-\rho$ and $\delta$ by $-\delta$ in
Eq.~\eqref{2/08/23-5} we find
\begin{equation}
\gamma(t)=\frac{1}{k_BT}\int_0^1t^{-\lambda}E_{{-\rho},1-\lambda}^{-\delta}
(-t^{{-\rho}})d\lambda \quad \text{and} \quad \left(\hat{\gamma}(s)
=\frac{1}{k_BT}\frac{s-1}{s\log s}(1+{s^{\rho}})^{\delta}\right),
\end{equation}
where $0<\rho,\delta<1$. The relaxation function $I(t)$ for $\omega\neq0$
reads
\begin{align}
\label{2/08/23-7}
I(t)&\sim_{t\rightarrow\infty}\mathcal{L}^{-1}\left\{\frac{s^{-1}}{\frac{1}{k_BT}\frac{s-1}{
\log s}(1+{s^{\rho}})^{\delta}+\omega^2};t\right\}\sim_{t\rightarrow\infty}\mathcal{L}^{-1}
\left\{\frac{s^{-1}}{\frac{1}{k_BT}\frac{s-1}{\log s}+\omega^2};t\right\}
\nonumber\\
&=\frac{1}{\omega^2}\mathcal{L}^{-1}\left\{\frac{s^{-1}}{1-\frac{1}{k_BT
\omega^2}\frac{1-s}{\log{s}}};t\right\}\sim_{t\rightarrow\infty}\frac{1}{\omega^2}\mathcal{L}^
{-1}\left\{\frac{1}{s}+\frac{1}{k_BT\omega^2}\frac{1}{s\log s};t\right\}
\nonumber\\
&=\frac{1}{\omega^2}\left[1+\frac{\nu(t)}{k_BT\omega^2}\right].
\end{align}
For $\omega=0$ the $t\to\infty$-asymptotic of $I(t)$ is equal to
\begin{align}
\label{2/08/23-72}
I(t)&=k_BT\mathcal{L}^{-1}\left\{\frac{s^{-1}}{\frac{s-1}{\log s}(1+{s^{\rho}})
^{\delta}};t\right\}\sim_{t\rightarrow\infty} k_BT\mathcal{L}^{-1}\left\{\frac{\log s}{s(s-1)};
t\right\}\nonumber\\
&=k_BT\big[C+\log t+e^tE_1(t)\big].
\end{align}
Note that the relevant MSD is the same as obtained for the force-free case by
using the distributed-order fractional derivative, i.e., Eq.~\eqref{MSD power
law noise distributed asymptotic} for $\alpha=1$.

\subsection{Langevin equation with distributed-order Volterra function}

We consider the GLE for which the friction kernel is based on the Volterra
functions $\nu(t,-\alpha)$ and $\mu(t,-\beta)$ for $0<\alpha,\beta<1$, see
~\ref{app3}. In the first case the friction memory kernel in $t$ and $s$
reads  
\begin{equation}
\label{3/08/23-1}
\gamma_1(t)=\frac{1}{k_BT}\int_0^1\nu(t,-\alpha)d\alpha \quad \text{and} \quad \hat{\gamma}_1(s)=\frac{1}{k_BT}\frac{s-1}{s\log^2s},
\end{equation}
thus satisfying condition \eqref{assumption}. For the GLE for the stochastic
harmonic oscillator ($\omega\neq0$) the long time limit ($t\to\infty$) of the
relaxation function $I_1(t)$ yields in the form
\begin{align}
\label{3/08/23-2}
I_1(t)&=\mathcal{L}^{-1}\left\{\frac{s^{-1}}{s^2+\frac{1}{k_BT}\frac{s-1}{\log^2
s}+\omega^2};t\right\}\sim_{t\rightarrow\infty}\frac{1}{\omega^2}\mathcal{L}^{-1}\left\{\frac{1}{s}
+\frac{1}{k_BT\omega^2}\frac{1}{s\log^2s};t\right\}\nonumber\\
&=\frac{1}{\omega^2}\left[1+\frac{\mu(t,1)}{k_BT\omega^2}\right].
\end{align}
In the force-free case ($\omega=0$) we obtain \cite[Eq.(45)]{KGorska}
\begin{align}
\label{3/08/23-3}
I_1(t)&=\mathcal{L}^{-1}\left\{\frac{s^{-1}}{s^2+\frac{1}{k_BT}\frac{s-1}{\log^2
s}};t\right\}\sim_{t\rightarrow\infty} k_BT\mathcal{L}^{-1}\left\{\frac{\log^2 s}{s(s-1)};t\right\}
\nonumber\\
&=k_BT\Big[-(C^2+\pi^2/6)e^t+2te^t\Phi_{1;1}^{\star,(1,1)}(-t,3,1)-2(C+\log t)
e^t E_1(t)\nonumber\\
&-(2C+\log t)(e^t+1)(\log t)-C^2+\pi^2/6\Big].
\end{align}
Here $\Phi_{1;1}^{\star,(1,1)}$ is the Hurwitz-Lerch function, see \cite {LiSr}

In the second case, the friction in $t$ and $s$ spaces turns out to be
\begin{equation}
\label{3/08/23-4}
\gamma_2(t)=\frac{1}{k_BT}\int_0^1\mu(t,-\beta)d\beta \quad \text{and} \quad
\hat{\gamma}_{2}(s)=\frac{1}{k_BT}\frac{(\log s)-1}{s(\log s)(\log \log s)}.
\end{equation}
Here we consider only the force-free case, i.e., $\omega=0$. In that case we
find the asumptotic of $I_2(t)$ at long times. Using that $\log\log s$ is a
slowly varying function, by Tauberian theorems we find that
\begin{align}
I_2(t)&\sim_{t\rightarrow\infty} k_BT\mathcal{L}^{-1}\left\{\frac{(\log s)(\log\log s)}{s(\log
s-1)};t\right\}=k_BT\mathcal{L}^{-1}\left\{\frac{\log\log s}{s(1-\frac{1}{\log
s})};t\right\}\nonumber\\ 
&\sim_{t\rightarrow\infty} k_BT\mathcal{L}^{-1}\left\{\frac{1}{s}{\log\log\frac{1}{s^{-1}}};t\right\}
=k_BT\log\log\frac{1}{t}.
\end{align}

\section{Summary}
\label{Sum}

We studied the GLE for an external harmonic potential and the free particle
limit for various cases of the friction memory kernel. In particular we
derived the associated MSD and DACF for the different chosen forms for the
friction memory kernel of Dirac delta, power-law, and distributed-order
forms. Various anomalous diffusive behaviors, such as subdiffusion,
superdiffusion, ultraslow diffusion, and strong anomaly, are observed.
Special attention was paid on distributed-order GLEs and distributed-order
diffusion-type equation. For different forms of the weight functions we
obtained ultraslow diffusion, strong anomaly, and other complex diffusive
behaviors.

It will be interesting to compare the results obtained here for the GLE to the
case for external noise \cite{klimo}, i.e., when the second FDT is not satisfied.
In both cases superstatistical and stochastic variations of the diffusion
coefficient and anomalous scaling exponent (Hurst exponent) have been analyzed 
recently \cite{diego1,chaos,elife,jakub,jakub1,wei}. Studying such concepts in
the frameworks developed here will significantly enlarge our current range of
stochastic models for disordered systems. We also note potential generalizations
with respect to subordinated generalized Langevin equation models such as those
studied in \cite{fox,yingjie}, fractional GLE~\cite{sandev3,sandev_jmp}, as well as presence of stochastic resetting~\cite{maj} in the system, including resetting in the memory kernel~\cite{fractal and fractional,TSbook}.

\section*{Acknowledgements}

RM and TS acknowledge financial support by the German Science Foundation (DFG,
Grant number ME~1535/12-1). TS is supported by the Alliance of International
Science Organizations (Project No.~ANSO-CR-PP-2022-05). TS is also supported
by the Alexander von Humboldt Foundation. KG and TP acknowledge financial support provided by the NCN Grant Preludium Bis 2 No. UMO-2020/39/O/ST2/01563.

\appendix

\section{Multinomial Prabhakar type functions ${\cal E}_{(\vec{\mu}),\beta}(t;
\vec{a})$}
\label{app1}

The multinomial Prabhakar function ${\cal E}_{(\vec{\mu}), \beta}(t; \vec{a})$ with $\mu_1 > \mu_2 > \ldots > \mu_m > 0$, see Ref.~\cite{EBazhlekova21}, is related to the multinomial ML function $E_{(\vec{\mu}), \beta}(-a_1 t^{\mu_1}, \ldots, -a_m t^{\mu_m})$ as follows
\begin{equation}\label{5/07/23-1}
{\cal E}_{(\vec{\mu}), \beta}(t; \vec{a}) = t^{\beta - 1} E_{(\vec{\mu}), \beta}(-a_1 t^{\mu_1}, \ldots, -a_m t^{\mu_m}).
\end{equation}
Its Laplace transform reads
\begin{equation}\label{5/07/23-2}
\mathcal{L}\{{\cal E}_{(\vec{\mu}), \beta}(t; \vec{a}); s\} = \frac{s^{\mu_1 - \beta}}{s^{\mu_1} + a_m s^{\mu_1-\mu_m}+\ldots + a_2 s^{\mu_1-\mu_2} + a_1}.
\end{equation}
Theorem~2.3 from Ref.~\cite{EBazhlekova21} implies that the following identities hold true:
\begin{align}\label{5/07/23-3a}
    \int_0^t \mathcal{E}_{(\vec{\mu}), \beta}(\xi; \vec{a}) d\xi & = \mathcal{E}_{(\vec{\mu}), \beta + 1}(t; \vec{a}), \\ \label{5/07/23-3b}
    \frac{d}{d t} \mathcal{E}_{(\vec{\mu}), \beta}(t; \vec{a}) & = \mathcal{E}_{(\vec{\mu}), \beta - 1}(t; \vec{a}), \quad \beta > 1.
\end{align}

The asymptotic behaviors of the multinomial Prabhakar function for $t \ll 1$ and $t\gg 1$ are equal to
\begin{align}\label{5/07/23-5a}
    \mathcal{E}_{(\vec{\mu}), \beta}(t; \vec{a}) &\sim_{t\rightarrow0} \frac{t^{\beta-1}}{\Gamma(\beta)} - \sum_{j=1}^{m} \frac{t^{\beta - 1 +\mu_j}}{\Gamma(\beta + \mu_j)},  \\ \label{5/07/23-5b}
    \mathcal{E}_{(\vec{\mu}), \beta}(t; \vec{a}) &\sim_{t\to\infty} a_2^{-1} t^{\beta - \mu_2 - 1} E_{\mu_1 - \mu_2, \beta - \mu_2}(-a_1 a_2^{-1} t^{\mu_1 - \mu_2}),
\end{align}
respectively. Eq.~\eqref{5/07/23-1} implies the series representation of ${\cal E}_{(\vec{\mu}), \beta}(s; \vec{a})$, that is
\begin{equation}\label{6/07/23-1}
    \mathcal{E}_{(\vec{\mu}), \beta}(t; \vec{a}) = \sum_{k=0}^{\infty} \sum_{k_1+\ldots+k_m = k \atop k_1, \ldots, k_m \geq 0} \frac{(-1)^k\, k!}{k_1!\ldots k_m!}\, \frac{\prod_{j=1}^m a_j^{k_j} t^{\beta - 1 + \sum_{j=0}^{m}\mu_j k_j}}{\Gamma(\beta + \sum_{j=1}^m \mu_j k_j)}
\end{equation}
from which it appears that we can represent the multinomial Prabhakar function by the sums of the three-parameter ML function \cite{Prabhakar}, i.e., 
\begin{equation}\label{three parameter ml}
E_{\mu, \beta}^{\delta}(-\lambda t^{\mu}) = \frac{1}{\Gamma(\delta)}\, \sum_{r=0}^{\infty} \frac{\Gamma(\delta + r)\, (-\lambda t^{\mu})^r}{r! \Gamma(\beta + \mu r)},
\end{equation}
whose the Laplace transform reads
\begin{equation}
\mathcal{L}\{t^{\beta-1} E_{\mu, \beta}^{\delta}(-\lambda t^{\mu}); s\}= \frac{s^{\mu\delta - \beta}}{(\lambda + s^{\mu})^\delta}
\end{equation}
For $\delta=1$ the three-parameter ML function becomes a two-parameter ML function, while for $\beta=\delta=1$ it becomes one parameter ML function. The asymptotic expansion of the three-parameter ML function follows from the expression~\cite{RGarra18, AGusti20}
\begin{equation}\label{GML_formula}
E_{\rho,\beta}^{\delta}(-z)=\frac{z^{-\delta}}{\Gamma(\delta)}\sum_{n=0}^{\infty}\frac{\Gamma(\delta+n)}{\Gamma(\beta-\rho(\delta+n))}\frac{(-z)^{-n}}{n!},
\end{equation} 
with $z>1$, and $0<\rho<2$.

For example, for $m=2$, we have
\begin{align}\label{7/07/23-1a}
    {\cal E}_{(\mu_1, \mu_2), \beta}(t; a_1, a_2) & = \sum_{j=0}^{\infty} (-a_2)^j t^{\beta + \mu_2 j - 1} E^{1+j}_{\mu_1, \beta + \mu_2 j}(-a_1 t^{\mu_1}) \\
    & = \sum_{j=0}^{\infty} (-a_1)^j t^{\beta + \mu_1 j - 1} E^{1+j}_{\mu_2, \beta + \mu_1 j}(-a_2 t^{\mu_2}). \label{7/07/23-1b}
\end{align}
Moreover, Eq.~\eqref{5/07/23-2} for $m = 2$ under condition $|a_1 s^{-\mu_1} + a_2 s^{-\mu_2}| >1$ gives
\begin{align}\label{7/07/23-2a}
    \mathcal{E}_{(\mu_1, \mu_2), \beta}(t; a_1, a_2) & = \sum_{r=0}^{\infty} \frac{(-1)^r}{a_1^{r+1}} t^{\beta -\mu_2(r+1) - 1} E_{\mu_1 - \mu_2, \beta - \mu_2(r+1)}^{r+1}\Big(-\frac{a_2}{a_1} t^{\mu_1 - \mu_2}\Big) \\ \label{7/07/23-2b}
    & = \sum_{r=0}^{\infty} \frac{(-1)^r}{a_2^{r+1}} t^{\beta -\mu_1(r+1) - 1} E_{\mu_2 - \mu_1, \beta - \mu_1(r+1)}^{r+1}\left(-\frac{a_1}{a_2} t^{\mu_2 - \mu_1}\right).
\end{align}

From Eq.~\eqref{6/07/23-1} for $m = 3$ after some laborious calculation can be derived, e.g., following formula which is used in our manuscript
\begin{align}\label{7/07/23-6}
    {\cal E}_{(\mu_1, \mu_2, \mu_3), \beta}(t; a_1, a_2, a_3) = \sum_{j=0}^{\infty} \frac{(-a_3)^j}{j!} \sum_{k=0}^{\infty} \frac{(-a_1)^k}{k!} (k+j)! t^{\beta-1 + \mu_1 k + \mu_3 j} E^{j+k+1}_{\mu_2, \beta + \mu_1 k +\mu_3 j}(-a_2 t^{\mu_2}).
\end{align}

\section{Tauberian theorems \cite{Feller}}\label{app2}

{\em 
If the Laplace transform pair $\hat{r}(s)$ of the function $r(t)$ behaves like
\begin{equation}\label{tauber5}
\hat{r}(s)\simeq s^{-\rho}L(s^{-1}), \quad s\to 0, \quad \rho > 0,
\end{equation}
where $L(t)$ is a slowly varying function at infinity, then $r(t)$ has the following asymptotic behavior
\cite{Feller}
\begin{equation}\label{tauber6}
r(t)\simeq \frac{1}{\Gamma(\rho)}t^{\rho-1}L(t), \quad t\to\infty.
\end{equation}
}
A slowly varying function at infinity means
that
\begin{equation}
\lim_{t\to\infty}\frac{L(at)}{L(t)}=1, \quad a>0.
\end{equation}
The Tauberian theorem works also for the opposite asymptotic, i.e., for $t\to 0$.

\section{The Volterra family functions}\label{app3}

Volterra's function is defined as follows \cite{AApelblat10, AApelblat13, Erdei}
\begin{equation}\label{eq:31082022-w10}
\mu(t, \beta, \alpha) = \frac{1}{\Gamma(1+\beta)} \int_{0}^{\infty} \frac{t^{u + \alpha}\, u^{\beta}}{\Gamma(u + \alpha + 1)} du, \qquad \Re(\beta) > -1 \quad \text{and} \quad t > 0,
\end{equation}
whose particular cases are
\begin{align*}
\alpha = \beta = 0: \qquad & \nu(t) = \mu(t, 0, 0), \\
\alpha \neq 0, \, \beta = 0: \qquad & \nu(t, \alpha) = \mu(t, 0, \alpha), \\
\alpha = 0, \, \beta\neq 0: \qquad & \mu(t, \beta) = \mu(t, \beta, 0).
\end{align*}
The Laplace transform of the Volterra's function $\mu(t, \beta, \alpha)$ is given by \cite{Erdei}
\begin{equation}\label{eq:31082022-w11}
\mathcal{L}\{\mu(t, \beta, \alpha); s\} = \frac{1}{s^{\alpha + 1} \log^{\beta + 1}s}.
\end{equation}

\section{Derivation of Eq. \eqref{12/07/23-1}}\label{app4}
Since,
\begin{align}\label{13/07/23-2}
\int_0^{\infty} e^{-st} t^a (\log^{\lambda - 1} t) dt &= \frac{\partial^{\lambda - 1}}{\partial a^{\lambda - 1}} \int_0^{\infty} e^{-st} t^a dt \nonumber\\ 
& = \frac{\partial^{\lambda - 1}}{\partial a^{\lambda - 1}}\left\{\frac{\Gamma(a+1)}{s^{a+1}}\right\} = I^{1-\lambda}_{0}\left\{\frac{\Gamma(a+1)}{s^{a+1}}\right\} \nonumber\\
& = \frac{\Gamma(a+1)}{\Gamma(1-\lambda)} \int^s_{0} (s-t)^{-\lambda} t^{-a-1} dt = \frac{\Gamma(a+1)\Gamma(-a)}{s^{\lambda + a}\Gamma(1 - \lambda - a)}, 
\end{align}
for $a=-1/2$, we get
\begin{equation}\label{13/07/23-3}
    \int_0^{\infty} e^{-st}t^{-1/2} (\log^{\lambda - 1}t) dt = \frac{\pi}{s^{\lambda - 1/2}\Gamma(3/2 - \lambda)}, \quad 0<\lambda<1.
\end{equation}
Then,
\begin{align}\label{13/07/23-4}
k_B T\, \hat{\gamma}(s) & =  \int_0^1 \Gamma(3/2 - \lambda) d\lambda \left[\int_0^{\infty} e^{-st} t^{-1/2} (\log^{\lambda - 1}t) dt\right] \nonumber\\
& =\pi\sqrt s\int_0^1\frac{d\lambda}{s^{\lambda}} = \pi\frac{s-1}{\sqrt s \log s}.
\end{align}

\end{document}